\documentclass[]{spie}

\usepackage[cmex10]{amsmath}
\usepackage{aas_macros}
\usepackage[numbers,sort&compress,nonamebreak]{natbib}
\usepackage{graphicx}
\usepackage[multidot]{grffile}
\usepackage{wrapfig}
\usepackage{floatrow}
\usepackage{caption}
\usepackage{subcaption}

\usepackage{multirow}
\usepackage{multicol}

\usepackage[backref=page,colorlinks=true,linkcolor=blue,urlcolor=blue,citecolor=blue]{hyperref}
\newcommand{\sref}[2]{\hyperref[#2]{#1\ref*{#2}}}

\newcommand{\mr}[2]{\multirow{#1}{*}{#2}}
\newcommand{\mc}[2]{\multicolumn{#1}{l}{#2}}
\newcommand{\pr}[2]{\mr{#1}{\parbox{1.5cm}{#2}}}

\newfloatcommand{capbtabbox}{table}[][\FBwidth]

\title{The REgolith X-Ray Imaging Spectrometer (REXIS) for OSIRIS-REx: Identifying Regional
Elemental Enrichment on Asteroids}

\author{
	Branden~Allen\supit{a}, 
	Jonathan~Grindlay\supit{a},
	Jaesub~Hong\supit{a},
	Richard~P.~Binzel\supit{b},
	Rebecca~Masterson\supit{c},
	Niraj~K.~Inamdar\supit{b,c},
	Mark~Chodas\supit{c},
	Matthew~W.~Smith\supit{c},
	Marshall~W.~Bautz\supit{d},
	Steven~E.~Kissel\supit{d},
	Joel~Villasenor\supit{d},
	Miruna~Oprescu\supit{a},
 	Nicholas~Induni\supit{a},
	\skiplinehalf
	\supit{a}Harvard College Observatory, 60 Garden Street, Cambridge, MA 02138, USA \\
	\supit{b}Department of Earth, Atmospheric, and Planetary Sciences, Massachusetts Institute of Technology, Cambridge, MA 02139, USA \\
	\supit{c}Space Systems Lab, Massachusetts Institute of Technology, Cambridge, MA 02139, USA \\
	\supit{d}Kavli Institute for Astrophysics and Space Research, Massachusetts Institute of Technology, Cambridge, MA 02139, USA \\
	}
\authorinfo{Further author information: (Send correspondence to Branden Allen; E-Mail:
\url{ballen@cfa.harvard.edu})}

\begin{document}
\maketitle
\begin{abstract}
The OSIRIS-REx Mission was selected under the NASA New Frontiers program and is scheduled for launch
in September of 2016 for a rendezvous with, and collection of a sample from the surface of asteroid
Bennu in 2019. 101955 Bennu (previously 1999 RQ$_\mathrm{36}$) is an Apollo (near-Earth) asteroid
originally discovered by the LINEAR project in 1999 which has since been classified as a potentially
hazardous near-Earth object.  The REgolith X-Ray Imaging Spectrometer (REXIS) was proposed jointly
by MIT and Harvard and was subsequently accepted as a student led instrument for the determination
of the elemental composition of the asteroid's surface as well as the surface distribution of select
elements through solar induced X-ray fluorescence.  REXIS consists of a detector plane that contains
4 X-ray CCDs integrated into a wide field coded aperture telescope with a focal length of 20 cm for
the detection of regions with enhanced abundance in key elements at 50 m scales.  Elemental surface
distributions of approximately 50-200 m scales can be detected using the instrument as a simple
collimator.  An overview of the observation strategy of the REXIS instrument and expected
performance are presented here.
\end{abstract}

\section{The OSIRIS-REx Mission}\label{section:mission}
The \textbf{O}rigins, \textbf{S}pectral \textbf{I}nterpretation, \textbf{R}esource
\textbf{I}dentification, \textbf{S}ecurity, \textbf{R}egolith, \textbf{Ex}plorer (OSIRIS-REx)
\cite{2012LPI....43.2491L,2012espc.conf..875B} is a multifaceted asteroid regolith sample return
mission chosen as part of NASA's New Frontiers Program for the characterization of a 101955 Bennu
(previously designated 1999 RQ$_\textrm{36}$).  OSIRIS-REx is equipped with
5 separate instruments: 
\begin{enumerate}
	\item The \textbf{O}SIRIS-REx \textbf{Cam}era \textbf{S}uite (OCAMS) which consists of three
	separate visible telescopes for the mapping and sample site selection on the asteroid at visible
	wavelengths \cite{2013LPICo1719.1690S}.
	\item The \textbf{O}SIRIS-REx \textbf{L}aser \textbf{A}ltimeter (OLA) which utilizes scanning
	LIDAR to provide high resolution topographical measurements of the asteroid surface
	\cite{2012LPI....43.1447D}. 
	\item Characterization of the asteroid spectra at visible and infrared wavelengths will be
	carried out using the \textbf{O}SIRIS-REx \textbf{V}isible and \textbf{IR S}pectrometer (OVIRS)
	\cite{2013LPICo1719.1100S,2012LPICo1683.1074R}.
	\item Thermal emission spectral maps will be provided by the \textbf{O}SIRIS-REx \textbf{T}hermal
	\textbf{E}mission \textbf{S}pectrometer (OTES).
	\item The \textbf{RE}golith \textbf{X}-Ray \textbf{I}maging \textbf{S}pectrometer (REXIS) will
	determine the global elemental abundances and search for anisotropies in the composition of Bennu.
\end{enumerate}
The spacecraft also carries the \textbf{T}ouch-\textbf{A}nd-\textbf{G}o \textbf{S}ample
\textbf{A}cquisition \textbf{M}echanism (TAGSAM) for the retrieval of no less than 60 g of regolith
from the surface of Bennu. 

The purpose and primary objectives of the OSIRIS-REx mission are the return of a regolith sample,
the global characterization of the asteroid's morphology and composition, measurement of the
properties of the sample site at sub-centimeter scales, and threat assessment with regard to
potential future impacts with the Earth. The threat assessment will be carried out by
preforming a field Yarkovsky test which will be used to refine future orbital projections.  Aside
from its power to extinguish life Bennu is also potentially important for advancing our
understanding of planet formation, in particular our understanding of the origin of organic material
which is of fundamental importance for the development of life on Earth (see
\sref{\S}{section:bennu}).  The measurements made in orbit will also be used to cross-calibrate
observations carried out from the vicinity of the Earth which should contribute to added precision
in the measurement of asteroid properties in general.

OSIRIS-REx is scheduled for launch in September of 2016, with a backup date scheduled one year
later.  After a gravity assist maneuver with the Earth in late 2017 OSRIS-REx will enter the
asteroid approach phase in late 2019 and carry out a preliminary survey for small natural satellites
which could pose a danger to the mission.  Once it is determined to be safe to proceed OSIRIS-REx
will settle into an orbit around Bennu and primary science operations will commence.  Observations
with the REXIS (see \sref{\S}{section:rexis}) will be performed in orbital phase B at a nominal
altitude of 720 m above the surface of the asteroid in order to assist with the selection of
potential sample retrieval sites by characterizing the global elemental abundance of Bennu and
searching for regions of anomalous composition.  After selection of a sample site OSIRIS-REx will
descend to the surface of Bennu and collect up to 2.2 kg of regolith.  OSIRIS-REx will depart the
asteroid in March of 2021 with its cargo which will land with the sample return capsule (SRC) at the
Utah Test and Training Range (UTTR) in September of 2023.

\section{Target: Near Earth Asteroid 101955 Bennu}\label{section:bennu}
\begin{wrapfigure}{r}{0.5\textwidth}
	\centering
	\includegraphics[width=0.99\textwidth]{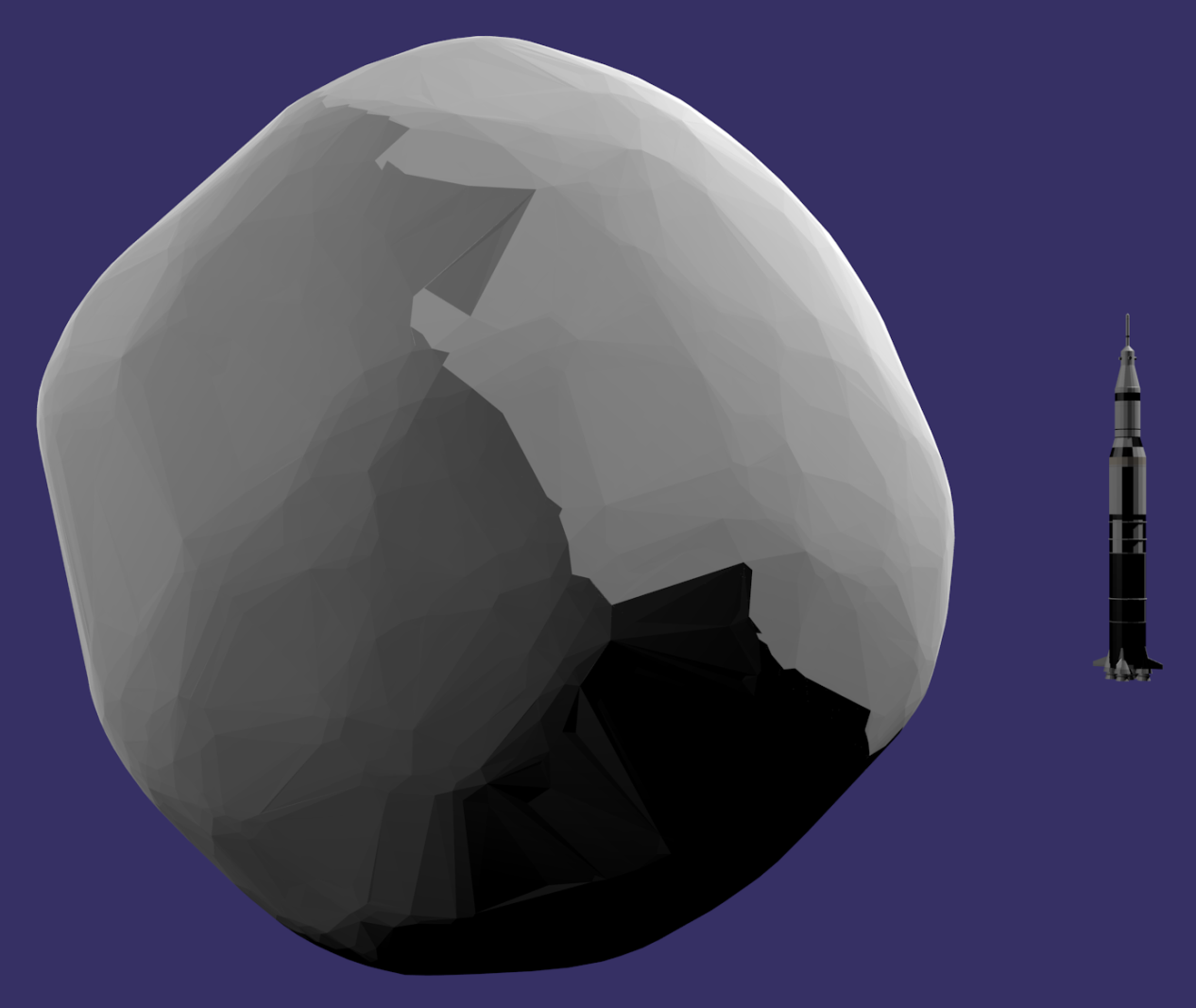}
	\caption{A comparison of 101955 Bennu with the Saturn 5 rocket.  Bennu has a mean radius of
	$246\pm10$ m, the Saturn 5 had a height of 110 m.}
	\label{figure:bennu}
\end{wrapfigure}

101955 Bennu (1999 RQ$_\mathrm{36}$), was selected by the OSIRIS-REx team for its accessibility,
relatively high impact probability, and it is a B-Type asteroid: a class that likely samples the
early chemistry of the solar system.  

Bennu was originally detected by the Lincoln Near-Earth Asteroid Program (LINEAR) in 1999
\cite{2000Icar..148...21S} and subsequently was determined to have the highest impact hazard rating
based on the Palermo Technical Scale (c.f. \cite{Chesley2002423}) with a cumulative value of
$-1.12$ and a maximum value of $-1.52$; for comparison the median values drawn from a sample of
$444$ asteroids currently tracked by the JPL Sentry System are
$P_\mathrm{cumulative}=-6.225\pm1.490$ and $P_\mathrm{maximum}=-6.545\pm1.474$ where the errors
stated here are the RMS values calculated over the sample \cite{Sentry_2013}.  The cumulative impact
probability with the Earth is $7.1\times10^{-4}$ with 8 potential impacts between 2169 and 2199.
The orbit of Bennu is in a psudo-resonance with the Earth's orbit and consequently makes close
passes at intervals of approximately 6 years; since its discovery ground based radar observations
have been performed during the close encounters of 1999, 2005 and 2011 allowing for detailed
characterization of the gross dimensions.  From these measurements it has been determined that Bennu
possesses a mean diameter of $492\pm20$ m with a total volume of $0.0623\pm0.006$ km$^3$ and a
surface area of $0.786\pm0.04$ km$^{2}$ (see \sref{figure }{figure:bennu}) \cite{Nolan2013629}.  The
shape of the asteroid is roughly spherical with a top-like appearance and a maximum diameter of
$565\pm10$ m.  From the same observations the sidereal rotation period was determined to be
$4.297\pm0.002$ hours and that the asteroid has a bulk density of $0.98\pm0.15$ g/cm$^3$
\cite{2013Icar..226..663H}.

Asteroids constitute the remaining building blocks of terrestrial planet formation and therefore
provide an important window into the conditions present during the formation of the Solar System.
More specifically primordial carbonaceous asteroids are a potential source of organic matter and
other volatile elements, such as sulfur, and may account for a large fraction of these elements on
the Earth.  Observations of Bennu are characterized by a low albedo of approximately $0.035\pm0.015$
and a lack of absorption bands characterizing other B-Type asteroids; the closest meteorite analog
was determined to be that of a C1 and/or CM1 Chondrite \cite{Clark2011462}.

\section{The Regolith X-Ray Imaging Spectrometer}\label{section:rexis}
\begin{table}
	\centering
	\small
	\begin{tabular}{llllll}\hline
		\mr{2}{Instrument}                                              & \mr{2}{Target}         & Obs. Start                            & Eng. Rng. [keV]                                   & Det. Area [cm$^2$]                                                 & \mr{2}{Notes}\\
		                                                                &                        & Obs. End                              & Eng. Res. [keV]                                   & FOV                                                                & \\\hline\hline
                                                                                               
		\mr{2}{Apollo 15 XRFS}                                          & \mr{2}{Moon}           & 30.7.1971                             & \mr{2}{\parbox{2.3cm}{1.5-5.5\\ --- }}            & \mr{2}{\parbox{2.3cm}{25.0\\30.0$^\circ$}}                         & 3 collimated prop. \\
		                                                                &                        & 4.8.1971                              &                                                   &                                                                    & counters \cite{1972NASSP.289......,1972LPSC....3.2157A} \\[1.5ex]
                                                                                               
		\mr{4}{Kaguya-XRS$^\bullet$}                                    & \mr{4}{Moon}           & \pr{4}{14.9.2007\\29.5.2009}          & \mr{4}{\parbox{2.3cm}{1.0-10.0\\0.180@5.9 keV}}    & \mr{4}{\parbox{2.3cm}{100.0\\12.0$^\circ$ $\times$ 12.0$^\circ$}}  & 4 collimators each\\
		                                                                &                        &                                       &                                                   &                                                                    & with 2 $\times$ 2 CCDs\\
		                                                                &                        &                                       &                                                   &                                                                    & (Rad. Damage)\\
		                                                                &                        &                                       &                                                   &                                                                    & \cite{2008LPI....39.1960O,2010SSRv..154....3K,2010TrSpT...7.Tk39O} \\[1.5ex]
                                                                                               
		\mr{3}{Hayabusa-XRS}                                            & \mr{3}{25143 Itokawa}  & \pr{3}{12.9.2005\\24.11.2005}         & \mr{3}{\parbox{2.3cm}{1.0-10.0\\0.14@1.5 keV}}    & \mr{3}{\parbox{2.3cm}{25.0\\3.5$^\circ$ $\times$ 3.5$^\circ$}}     & 1 collimator\\
		                                                                &                        &                                       &                                                   &                                                                    & with 2 $\times$ 2 CCDs\\
		                                                                &                        &                                       &                                                   &                                                                    & \cite{2000AdSpR..25..345O,2002AdSpR..29.1237O,2006Sci...312.1338O} \\[1.5ex]
                                                                                               
		\mr{3}{Chandrayaan-1 C1XS}                                      & \mr{3}{Moon}           & \pr{3}{11.2008\\8.2009}               & \pr{3}{0.8-20.0\\0.2}                             & \mr{3}{\parbox{2.3cm}{24.0\\12.0$^\circ$}}                         & 24 SCDs in \\
		                                                                &                        &                                       &                                                   &                                                                    & 3 collimators \\
		                                                                &                        &                                       &                                                   &                                                                    & \cite{Howe2009735,Narendranath201153} \\[1.5ex]
                                                                                               
		\mr{2}{SMART-1 D-CIXS}                                          & \mr{2}{Moon}           & \pr{2}{3.2005\\3.2005}                & \pr{2}{1.0-20.0\\0.2}                             & \pr{2}{12.0\\12$^\circ$}                                           & 24 SCDs in \\
		                                                                &                        &                                       &                                                   &                                                                    & 3 collimators \cite{Grande2003427} \\[1.5ex]
                                                                                               
		\mr{2}{NEAR-XRS}                                                & \mr{2}{443 Eros}       & \pr{2}{4.5.2000\\10.2.2001}           & \mr{2}{\parbox{2.3cm}{1.5-5.5\\0.83@5.9 keV}}     & \pr{2}{25.0\\5$^\circ$}                                            & 3 collimated prop.\\
		                                                                &                        &                                       &                                                   &                                                                    & counters \cite{2001MPS...36.1673N} \\[1.5ex]
                                                                                               
		Mercury Messenger                                               & \mr{2}{Mercury}        & \pr{2}{2011\\Present}                 & \mr{2}{\parbox{2.3cm}{1.0-10\\0.88@5.9 keV}}      & \mr{2}{\parbox{2.3cm}{10.0\\12.0$^\circ$}}                         & 3 collimated prop.\\
		XRS                                                             &                        &                                       &                                                   &                                                                    & counters \cite{2007SSRv..131..393S,2011Sci...333.1847N} \\[1.5ex]
                                                                                               
		\pr{3}{Bepicolombo\\MIXS-C$^\dagger$}                           & \mr{3}{Mercury}        & \pr{3}{8.2019\\8.2020}                & \mr{3}{\parbox{2.3cm}{0.5-10.0\\0.128@5.9 keV}}   & \mr{3}{\parbox{2.3cm}{3.68\\10.4$^\circ$}}                         & Single APS in a\\
		                                                                &                        &                                       &                                                   &                                                                    & single collimator\\
		                                                                &                        &                                       &                                                   &                                                                    & \cite{2010PSS...58...79F,2010NIMPA.624..540T} \\[1.5ex]
                                                                                            
		\pr{3}{Bepicolombo\\MIXS-T$^{\dagger\ast}$}                     & \mr{3}{Mercury}        & \pr{3}{8.2019\\8.2020}                & \mr{3}{\parbox{2.3cm}{0.5-10.0\\0.128@5.9 keV}}   & \mr{3}{\parbox{2.3cm}{3.68\\1.1$^\circ$}}                          & APS w/Focusing optic\\
		                                                                &                        &                                       &                                                   &                                                                    & 6$^\prime$ angular res.\\
		                                                                &                        &                                       &                                                   &                                                                    & \cite{2010PSS...58...79F,2010NIMPA.624..540T} \\[1.5ex]
                                                            
		\mr{3}{\parbox{2.3cm}{OSIRIS-REx\\REXIS$^{\dagger\ddagger}$}}   & \mr{3}{101955 Bennu}   & \pr{3}{12.2019\\3.2020}               & \mr{3}{\parbox{2.3cm}{0.3-10.0\\0.130@5.9keV}}    & \mr{3}{\parbox{2.3cm}{24.2\\30.0$^\circ$}}                         & 2 $\times$ 2 CCDs\\
		                                                                &                        &                                       &                                                   &                                                                    & Coded-aperture tel.\\
		                                                                &                        &                                       &                                                   &                                                                    & $26.2^\prime$ angular res.\\\hline
%		\mc{2}{$^\diamond$Solar Monitor Failure}                                                 &                                       &                                                   &                                                                    & \\                              
		\mc{2}{$^\bullet$Radiation Damage}                                                       &                                       &                                                   &                                                                    & \\                              
		\mc{2}{$^\dagger$Future Missions}                                                        &                                       &                                                   &                                                                    & \\                                
		\mc{2}{$^\ddagger$Coded-Aperture Telescope}                                              &                                       &                                                   &                                                                    & \\                              
		\mc{2}{$^\ast$Focusing Telescope}                                                        &                                       &                                                   &                                                                    & \\                                
	\end{tabular}
	\caption{A list of previous remote sensing X-ray fluorescence experiments flown or scheduled to
	be flown for the characterization of elemental abundances on airless bodies within the solar
	system.}
	\label{table:instruments}
\end{table}

The \textbf{RE}golith \textbf{X}-Ray \textbf{I}maging \textbf{S}pectrometer (REXIS) (shown in
\sref{figure }{figure:rexis}) was conceived as a student led project whose primary goal is the
education of science and engineering students who will participate in the development of flight
hardware in future space missions.  Additionally REXIS also augments the observation capabilities of
the OSIRIS-REx mission at the high end of the electromagnetic spectrum which will enable
characterization of the asteroid elemental abundances from a global scale down to 50 m, a capability
unique to REXIS among instruments of this type that have previously flown (c.f. \sref{table
}{table:instruments}).

\begin{figure}
	\begin{floatrow}
		\ffigbox{%
			\centering
			\includegraphics[width=0.49\textwidth]{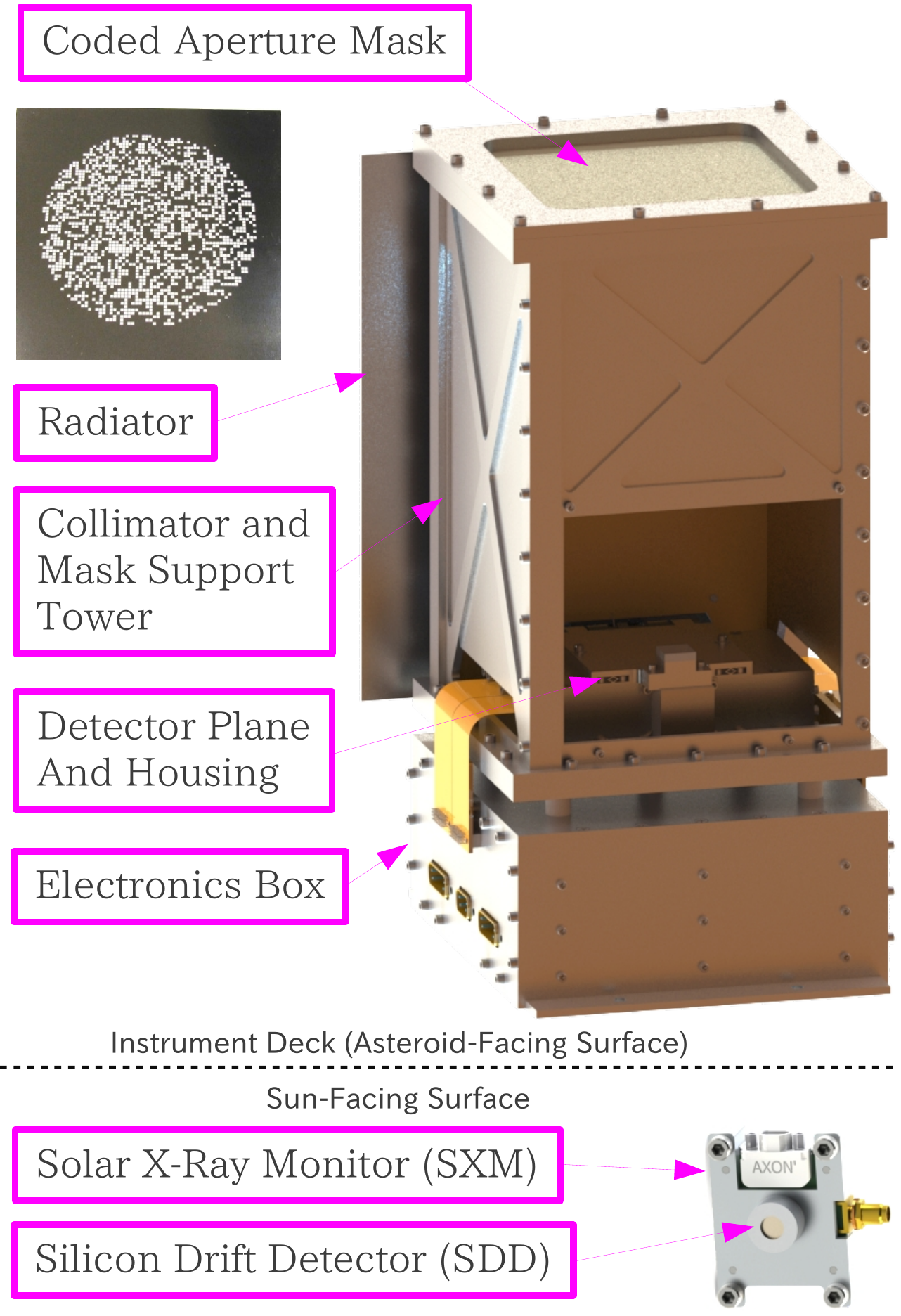}
		}{%
			\caption{The REXIS instrument consists of the main detector and collimator assembly and the
			electronics box which are physically connected by 5 thermal isolation standoffs. The mating
			interface to the OSIRIS-REx spacecraft are located on the underside of the electronics box.
			Shown to the upper left is the prototype coded aperture mask which makes detection of
			elemental abundance enhancements at 50 m scales possible. The solar X-ray monitor (SXM) is
			mounted on the sun-facing side of the spacecraft separately from the primary telescope.}
			\label{figure:rexis}
		}
		\capbtabbox{%
			\centering\small
			\begin{tabular}{ll}\hline
				\mc{2}{\textbf{Physical Parameters}}                                                                          \\\hline\hline
				Total Mass (CBE)                            & 4.4 kg                                                          \\
				Total Power (CBE)                           & 10.8 W                                                          \\
				Focal Length                                & 20 cm                                                           \\
				Detector Plane                              & 2 $\times$ 2 CCDs                                               \\
				Active Area                                 & 24.159 cm$^2$                                                   \\
				FOV                                         & 0.18 sr (30$^\circ$ circ.) (FWZI)                               \\
				Angular Resolution                          & $26.2^\prime$ (5.6 m @ 730 m)                                   \\\hline
				\mc{2}{\textbf{CCD Parameters}}                                                                               \\\hline\hline
				Type                                        & MIT-LL CCID-41                                                  \\
				Energy Resolution                           & 130 eV @ 5.9 keV                                                \\
				Energy Range                                & 0.3-10.0 keV (QE$\geq$0.3)                                      \\
				Pixels                                      & 1024 $\times$ 1024                                              \\
				Pixel Dimensions                            & 24 $\mu$m $\times$ 24 $\mu$m                                    \\
				Active Area                                 & 6.03880 cm$^2$                                                  \\
				Super-Pixel Dimensions                      & 0.192 mm $\times$ 0.192 mm                                      \\
				Depletion Depth                             & 45 $\mu$m                                                       \\
				Optical Blocking                            & 220 nm Direct Deposited Al                                      \\
				Operating Temperature                       & $\leq-$60$^\circ$C                                              \\\hline
				\mc{2}{\textbf{Mask Parameters}}                                                                              \\\hline\hline
				Thickness                                   & 100 $\mu$m                                                      \\
				Composition                                 & ASI-301 Stainless Steel                                         \\
				Pattern Diameter                            & 98.304 mm (64 Pixels)                                           \\
				Open Hole Fraction                          & 0.5                                                             \\
				Pixel Pitch                                 & 1.536 mm                                                        \\
				Support Grid Width                          & 100 $\mu$m                                                      \\\hline
				\mc{2}{\textbf{Solar X-Ray Monitor (SXM)}}                                                                    \\\hline\hline
				Detector                                    & Amptek XR-100 SDD                                               \\
				Active Area                                 & 5 mm $\times$ 5 mm                                              \\
				Energy Range                                & 1-20 keV                                                        \\
				Energy Resolution                           & 125 eV @ 5.9 keV                                                \\
				Depletion Depth                             & 500 $\mu$m                                                      \\      
				Optical Blocking                            & 0.5 mil Be Window                                               \\
				Operating Temperature                       & $\leq$0$^\circ$C                                                \\\hline
			\end{tabular}
		}{%
			\caption{A summary of the critical instrument parameters for REXIS, the CCDs, the coded
			aperture mask and the solar X-ray monitor (SXM) shown to the left in \sref{figure
			}{figure:rexis}.  The values given for the mass and power consumption of the instrument are
			the current best estimates (CBE).}
			\label{table:rexis}
		}
	\end{floatrow}
\end{figure}

REXIS is designed to observe induced X-ray fluorescence lines emitted from the asteroid surface that
arise as a result of exposure to solar X-rays as well as the cosmic X-ray background (CXB).  A
number of previous missions have employed this technique for the observation of airless bodies
throughout the Solar System beginning with the Apollo 15 X-ray fluorescence experiment for the
observation of the moon \cite{1972LPSC....3.2157A}.  More recent observations of the Moon have been
carried out by the D-CIXS mapping spectrometer on SMART-1 \cite{Grande2003427}, the C1XS on
Chandrayaan-1 \cite{Howe2009735,Narendranath201153,Crawford2009725}, and the XRS aboard Kaguya
\cite{2008LPI....39.1960O,2010SSRv..154....3K}.  Additionally similar observations of two separate
asteroids 433 Eros and 25143 Itokawa have been carried out by the Hayabusa-XRS
\cite{2002AdSpR..29.1237O,2006Sci...312.1338O} and NEAR-XRS \cite{2001MPS...36.1673N} respectively,
and the Mercury Messenger mission is currently in the process of mapping the elemental abundances on
the surface of Mercury \cite{2011Sci...333.1847N}.  To date the instruments that have flown have all
employed a collimator based instrument tuned to the specific orbital and target parameters imposed
by the constraints of the individual missions (see \sref{table }{table:instruments}).  This allows
for some reconstruction of the elemental abundances as a function of position on the object under
study, however the angular resolution is defined by the size of the collimator and is therefore
rather coarse compared to what is achievable utilizing focusing X-ray optics or a coded-aperture
instrument.

\begin{wrapfigure}{r}{0.5\textwidth}
	\centering
	\includegraphics[width=0.99\textwidth]{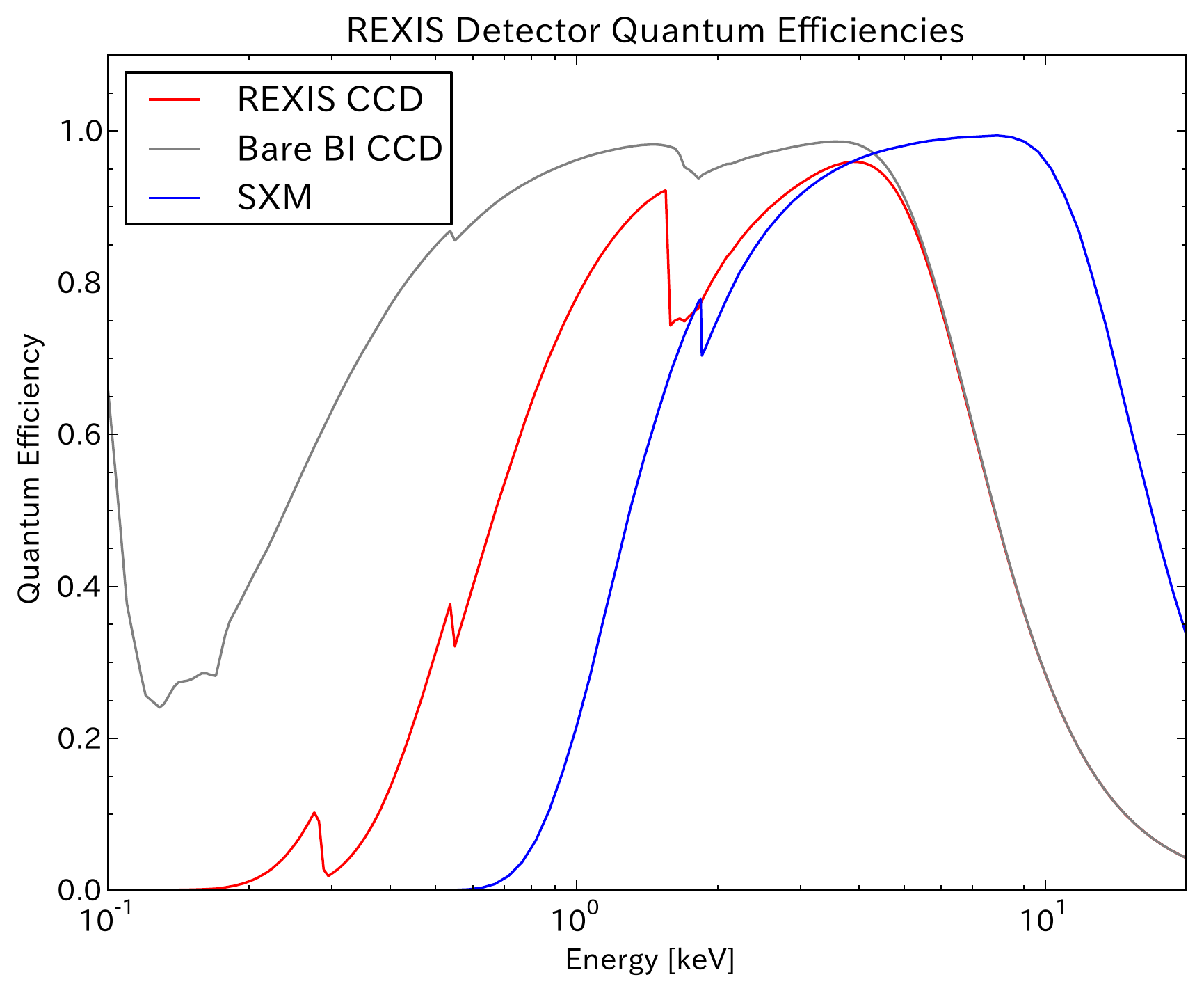}
	\caption{The on-axis quantum efficiency curve of the bare MIT Lincoln Laboratory CCID-41 and the
	predicted performance of the REXIS flight CCDs after application of the 220 nm thick Al optical
	blocking filter and taking into account the effects of molecular surface contamination between
	0.1 and 20 keV.  The solar X-ray monitor (SXM) quantum efficiency including the 0.5 mil thick Be optical
	blocking window is also shown.}
	\label{figure:QE}
\end{wrapfigure}

A variety of detector planes have been used in conjunction with these missions.  To date XRF
experiments flown by National Aeronautics and Space Administration (NASA): the Apollo 15 XRFS,
NEAR-XRS, and the Mercury Messenger XRS have been limited to the use of essentially the same
instrument consisting of 3 gas proportional counters (GPC) where one GPC is outfitted with an Al
filter, another is outfitted with a Mg filter, and the third is simply equipped with a Be optical
blocking filter in order to deconvolve the Al-K, Mg-K, and Si-K emission lines in spite of the poor
energy resolution (880 eV @ 6.5 keV) of these detectors.  More recent efforts supported by the
Japanese Aerospace Exploration Agency (JAXA) have focused on the use of 4 $\times$ 4 arrays of X-ray
CCDs integrated into collimators and have the advantage of substantially improved energy resolution
140 eV @ 1 keV.  The primary drawback of this method is the susceptibility of CCDs to radiation
damage, in particular from low energy proton induced charge transfer inefficiencies (CTI), which
prevented the success of one instrument: the Kaguya-XRS.  Until recently the efforts of the European
Space Agency (ESA) were focused on the use of swept charge devices (SCD) which are non-imaging
silicon based detectors similar to a CCD but with an altered electrode layout optimized for rapid
readout that have energy resolutions comparable to CCDs.  The ESA Bepicolombo mission to Mercury,
set to begin observations in August of 2019, will employ two XRF instruments: a telescope consisting
of a collimator and a Depleted P-Channel Field Effect Transistor (DEPFET) array detector plane
(MIXS-C) and an identical detector plane outfitted with a focusing optic (MIXS-T)
\cite{2010NIMPA.624..540T,2010PSS...58...79F}.

The primary REXIS instrument consists of a detector plane comprised of a 2 $\times$ 2 array of back
illuminated (BI) CCDs, MIT Lincoln Laboratory (LL) CCID-41, identical to those used in the X-Ray
Imaging Spectrometer (XIS) \cite{2007PASJ...59S..23K} on Suzaku \cite{2007PASJ...59S...1M}, a joint
astrophysics mission between NASA and JAXA, which have quantum efficiencies greater than 0.75
between 0.4 and 6.0 keV (see \sref{figure }{figure:QE})
\cite{2004SPIE.5501..111B,2006SPIE.6266E..79B}.  The active area of the CCD is composed of a 1024
$\times$ 1024 array of pixels with a pitch of 24 $\mu$m.  There is a gap of 38 detector pixels
(0.912 mm) between the active areas of the individual CCDs to ensure the safe assembly of the
detector plane which are aligned in order to preserve the pixel pitch across the entire detector
plane.  A optical blocking filter consisting of a 220 nm aluminium layer deposited directly on the
CCD is present to mitigate the undesirable background signal that would otherwise be induced by
optical/UV photons reflected from the surface of the asteroid and onto the detector plane.  The
detector plane is housed in a thermally isolated, shielded box which itself is located inside of the
instrument main truss that supports a coded aperture mask located 20 cm above the detector plane.
The detector plane is passively cooled to less than $-60^\circ$C by means of a radiator attached to
one surface of the main truss and is oriented toward deep space during science operations.  This
entire assembly sits atop and is affixed with low thermal conductivity standoffs to the REXIS
electronics box which contains the instrument's support electronics.  The mask itself is
manufactured from a single piece of 100 $\mu$m thick ASI-301 stainless steel with a random pattern
of chemically etched open holes.  The pitch of the individual mask elements is 1.536 mm with an open
hole fraction of 0.5 and a support grid running between the individual elements of 100 $\mu$m.  The
support grid reduces the throughput to 0.437. In order to guard against radiation induced CTI and
preserve the spectral resolution of the detector plane the REXIS instrument will be outfitted with a
radiation cover at the aperture which will remain closed until observations commence.  The detector
plane is also equipped with a number of $^{55}$Fe calibration sources which are arranged to enable
the monitoring of a subset of detector pixels at the beginning and end of the CCD readout chain
throughout observations of Bennu.  The underside of the radiation cover, mounted above the coded
aperture mask, also contains an additional series $^{55}$Fe sources which will be used to calibrate
and monitor all pixels in the detector plane prior to the opening of the cover for science
observations.

The solar X-ray spectrum is highly variable characterized by episodic flares and outbursts typically
on 600 s timescale with longer durations on the order of days possible; on longer time scales
changes in flux of up to 3 orders of magnitude is also possible.  For this reason REXIS, as well as
all other XRF experiments of this type, must be equipped with an additional sensor for the
measurement of the local solar X-ray spectrum in order to properly interpret the data.  To this end
REXIS includes a separate solar X-ray monitor (SXM) which is attached to the sun-facing side of the
spacecraft and interfaces directly with the REXIS main detector electronics.  The SXM is constructed
using a commercially available silicon drift detector (SDD) manufactured by Amptek read out by
custom designed electronics originally commissioned for use in the Neutron Star Interior Composition
ExploreR (NICER) \cite{2012SPIE.8443E..13G,2012SPIE.8453E..18P} that have been optimized for use
with the REXIS instrument.  The SSD has an active area of 25 mm$^2$ and is housed behind a 0.5 mil
thick optical blocking filter composed of Be.  In order to maintain a nominal operating temperature
less than 0$^\circ$C under direct exposure from the sun a thermoelectric cooler is located directly
beneath the SDD.  The fully assembled SXM will have a quantum efficiency greater than 0.1 between
1.1 and 30.0 keV with an energy resolution of 125 eV at 5.9 keV (see \sref{figure }{figure:QE})
\cite{6154116,2012SPIE.8453E..18P}.

The primary REXIS instrument in conjunction with the SXM will carry out observations of the asteroid
Bennu and measure the asteroid elemental composition (see
\sref{\S}{section:asteroid_identification}) and conduct a search for regions of enhanced abundance
as described in \sref{\S}{section:imaging}.

\section{Observation Strategy}\label{subsection:observation}
After an initial cruise phase and survey (see \sref{\S}{section:mission}) the OSIRIS-REx spacecraft
will enter a roughly circular polar terminator orbit with a radius of 1 km or $730\pm20$ m above the
surface of the asteroid with an orbital period of 27 hours.  11 hours of observation each day are
currently allocated for nadir pointing observations during which time the REXIS field of view will
be centered on the asteroid.  The position of the REXIS-SXM has been optimized for this observation
period so that the sun remains in the SXM field of view to enable continuous monitoring of the solar
X-ray spectrum throughout science operations.  Concordantly a radiator affixed to the main
collimator and mask support structure for passive cooling of the CCDs to less than -60$^\circ$C has
been mounted so that it is oriented into deep space during these observations as well.

%CHECK FRAME TRANSFER TIME 6.7 $\mu$s from XIS paper
The CCDs in the detector plane are read out with an integration time of 4 seconds per frame, and a
frame transfer time of 10 ms.  The individual frames are processed on board where the individual
events from each frame are extracted, graded and placed in an event list with a time tag before
being staged on board the OSIRIS-REx spacecraft for downlink.  The 4 second timing resolution of the
detector plane combined with the asteroid rotation period and nominal orbit of OSIRIS-REx during
science operations set the lower bounds for the angular resolution of the REXIS instrument at
3.6$^\prime$, equivalent to a 28 cm spatial resolution on the surface of the asteroid from an
altitude of 730 m.  The maximum spatial resolution of the detector plane is fixed by the CCD pixel
size of 24 $\mu$m $\times$ 24 $\mu$m, however, since this level of fidelity is not required pixels
are re-binned in the on board data processing to super pixels consisting of a 8 $\times$ 8 array of
native pixels with a pitch of 0.192 $\mu$m.  This is done primarily to reduce the data volume of the
instrument since the full spatial resolution of the detector plane is not required to achieve
REXIS's science objectives

Science observations will commence just after the next solar minimum during which time a very low
incident solar X-ray flux is expected at the level defined by the Geostationary Operational
Environmental Satellite (GOES) mission A-state, i.e $10^{-8}$ W/m$^2$ in the 1-8 \AA\ band solar
X-ray monitor \cite{doi:10.1117/12.254082}.  In order to mitigate this the REXIS field of view has
been maximized to enable the collection of as many X-ray events as possible while maintaining
minimal exposure to the cosmic X-ray background which is largely obscured by the asteroid in the
REXIS field of view.  This comes at the expense of reduced spatial resolution on the surface of the
asteroid for a collimator based analysis.  In an effort to measure elemental abundance with high
spatial resolution ($26.2^\prime$) a coded aperture mask with a open hole fraction of 0.5 and pixel
pitch of 1.536 mm will be affixed at a distance (focal length) of 20 cm from the surface of the
detector plane but reduces the total aperture throughput (see \sref{figure }{figure:rexis}) by a
factor of two.  Coded aperture instruments have been used in a variety of astrophysics missions, in
particular wide-field hard X-ray monitors sensitive at energies between approximately 15 and 200 keV
such as the Swift Burst Alert Telescope (BAT) \cite{2005SSRv..120..143B} and the Imager on-board
INTEGRAL (IBIS) \cite{2003A&A...411L.131U} for the localization, long-term monitoring and
characterization of astrophysical point sources.  This method was pioneered as a way to image hard
X-rays with energies above approximately 10 keV where the use of focusing optics was not possible
until recently.  In the case of REXIS, which is sensitive between 0.4 and 10 keV, the use of a
coded-aperture mask was born primarily out of budget, mass constraints and FOV.

REXIS will be the first application of the coded aperture imaging technique to extended sources.
Since the surface being imaged provides the very background against which enhanced emission features
are sought, the technique is less sensitive than for point sources (for which it has been
traditionally used). However, combined with "collimator mode" imaging (see
\sref{\S}{section:recon}), the coded aperture technique will reveal localized features above some
minimum enhancement value.  A coded aperture imager operates as a pinhole camera, where additional
pinholes have been added in order to increase the total throughput of the instrument.  The
reconstruction of images from coded aperture instruments proceeds directly from analysis of the
X-ray event list by selecting events with the energies within the band of interest then generating
detector plane images using the spatial information recorded in the individual CCDs for each
exposure. The reconstruction of the local field for each individual exposure is achieved by
convolving the detector plane image with the mask image as described in \cite{1978ApOpt..17..337F}
and in \sref{\S}{section:imaging}; this is analogous to a search for the presence of significant
projections of the mask pattern on the detector plane from all possible angles in X-rays.  A final
asteroid image is achieved by co-addition of the individual reconstructed sky images reprojected
onto the surface asteroid surface (see \sref{\S}{section:imaging} for details).  Finally the angular
resolution of a coded aperture instrument is defined by 
\begin{equation}
	\delta\phi=\tan^{-1}\left(\frac{\sqrt{p_m^2+p_d^2}}{f}\right)
\end{equation}
where f is the separation between the detector plane and the coded-aperture mask (i.e. focal
length), $p_m$ and $p_d$ is the pitches of the mask and detector pixels respectively.  The angular
resolution of the REXIS instrument is $26.2^\prime$, about a factor of 7.3 below the maximum
achievable based on the CCD integration time and the motion of the spacecraft and asteroid.  This
is equivalent to a spatial resolution on the asteroid surface of 5.6 m at an altitude of 730 m.

\section{Asteroid X-Ray Emission Modeling}\label{section:asteroid_spectrum}
\begin{figure}
	\centering
	\begin{subfigure}{0.49\textwidth}
		\centering
		\includegraphics[width=\textwidth]{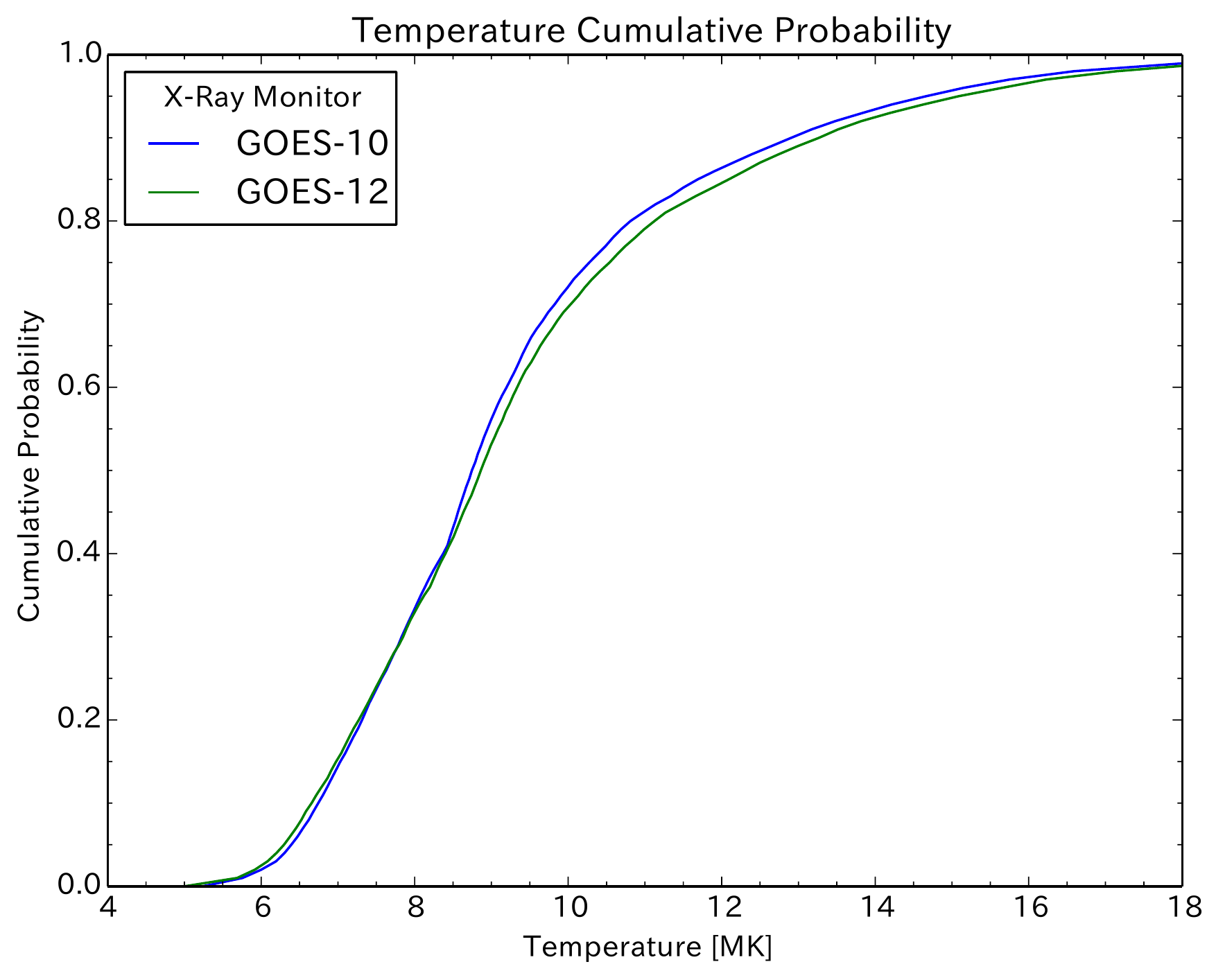}
	\end{subfigure}
	\begin{subfigure}{0.49\textwidth}
		\centering
		\includegraphics[width=\textwidth]{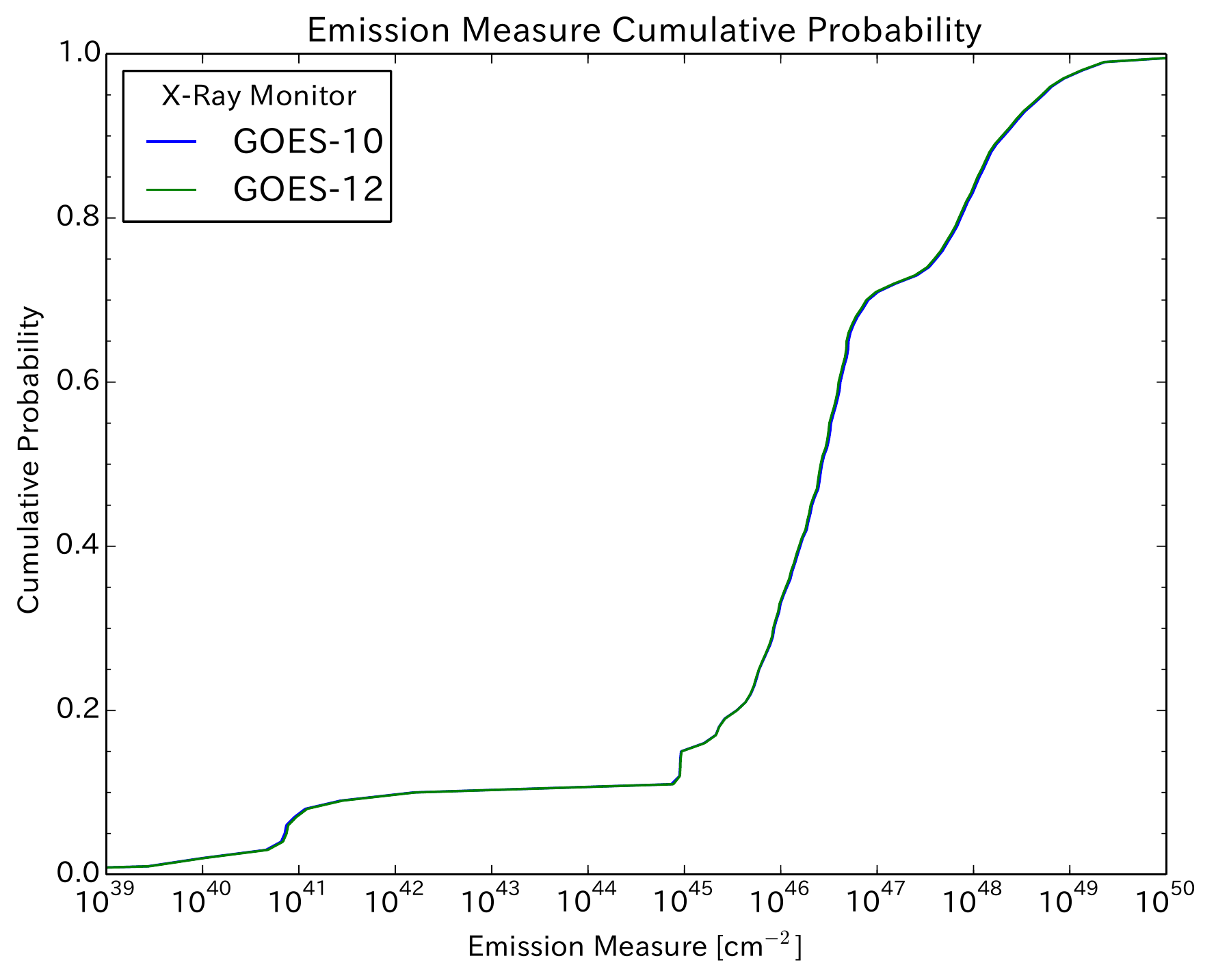}
	\end{subfigure}
	\caption{The cumulative probability distribution of the measured temperature and emission
	measures of the sun using data from the GOES-10, and 12 X-ray monitors during the time around
	the previous solar minimum between January 1, 2005 and January 1, 2007: the closest analog for
	the REXIS observing period.  A cut on the data has been implemented requiring agreement between
	the monitor data of 5\% in order to exclude instrumental and environmental effects.}
	\label{figure:goes_em_temp}
\end{figure}

To evaluate the performance of REXIS spectral modeling was carried out primarily using the
composition of the nearest meteorite match to the observed spectrum of 101955 Bennu (see
\sref{\S}{section:bennu}): that of a C1 chondrite.  The single interaction model presented in
\cite{1997JGR...10216361C} was adopted and cross checked with an independent GEANT4.3 simulation.
The current calculation assumes a smooth spherical asteroid with a uniform density of 1 g/cm$^3$
(see \sref{\S}{section:bennu}).

The input solar spectrum for baseline simulation was chosen using data from the GOES X-ray monitors
\cite{doi:10.1117/12.254082} in order to estimate the expected temperature and emission measures
around the depth of solar minimum, during which REXIS will carry out observations.  Using data
extracted for the GOES-10 and GOES-12 X-ray monitors averaged over 5 minute intervals and the method
introduced in \cite{2005SoPh..227..231W} emission measures and temperatures from the time around the
previous solar minimum were derived.  Taking a simplistic, conservative approach derived emission
measures and temperatures from the GOES-10 and GOES-12 satellites were compared and data points
found to differ by more than 5\% were excluded from consideration.  Cumulative probability
distributions were calculated for both (see \sref{figure }{figure:goes_em_temp}) and a worst-case
scenario temperature 4 MK and an emission measure of $10^{44}$ cm$^{-3}$ were adopted and used to
generate single temperature models using CHIANTI 7.1 \cite{1997A&AS..125..149D,2013ApJ...763...86L}.
Additionally these ``worst-case'' values were compared to results from Reuven Ramatti High Energy
Spectroscopic Imager (RHESSI) which place constraints on the solar X-ray spectrum at solar minimum
during periods where the GOES 1-8 \AA\ monitors register solar activity below $10^{-8}$ W/m$^{2}$,
i.e. below the A-state, and found to in be good agreement with this choice of parameters
\cite{2007ApJ...659L..77H}. In addition coronal plasmas at higher temperature states were also
considered for the imaging simulation (see \sref{\S}{section:imaging}).

\begin{wraptable}{l}{0.4\textwidth}
	\centering
	\begin{tabular}{ll}\hline
		Line              & Min. Det. Time \\\hline\hline
		Fe-L              & 136 sec.\\
		Mg-K              & 63 sec.\\
		Al-K              & 7.67 hrs.\\
		Si-K              & 183 sec.\\
		S-K$\alpha$       & 5.73 days\\\hline
		\mc{2}{Internal bkg. lines not included}
	\end{tabular}
	\caption{The estimated minimum detection times for key elements for an asteroid with the
	composition of a C1-Chondrite at a significance of 6$\sigma$.}
	\label{table:detection_times}
\end{wraptable}

After calculation of the solar spectrum incident on the asteroid, the induced X-ray fluorescence
(XRF) spectrum is derived for a simple spherical asteroid by numerical integration.  To accomplish
this a map of the asteroid is divided into equal area bins using an AITOFF-Hammer projection,
identical to that used in the imaging routines described in \sref{\S}{section:imaging}.  The
incidence angle and re-emission angle to the position of the OSIRIS-REx spacecraft are derived for
each bin then, using the prescription in \cite{1997JGR...10216361C}, the coherent and incoherent
scattering intensities as well as the intensity of the individual fluorescence lines are calculated.
The incoming flux is then integrated over all bins visible to the REXIS instrument weighted by the
response function of the coded-aperture mask shown in \sref{figure }{figure:open_fraction} giving
the total solar induced XRF spectrum.

In addition the XRF spectrum induced by the cosmic X-ray background (CXB) is calculated in a similar
fashion assuming a completely uniform CXB.  Under this assumption each bin on the asteroid surface
receives the same incident flux from a 2$\pi$ radian region on the sky.  Again following
\cite{1997JGR...10216361C} but integrating over all possible incidence angles each component of the
XRF spectrum is calculated for a fixed spacecraft position giving the total CXB induced XRF
spectrum.  Combination of these two components give the total predicted count rate for each line of
interest during low solar activity incident on the surface of the CCDs.  For this time period the
CXB induced XRF emission becomes important for elements at energies above 3 to 4 keV.  For this
calculation particle induced X-ray emission (PIXE) is not yet included but will be incorporated into
future models.  

Next the attenuation due to the presence of the 220 nm thick Al optical blocking filter and due to
the presence of organic contaminants on the surface of the CCD is applied along with the quantum
efficiency of the detectors as a function of energy (see \sref{figure }{figure:QE}) to produce a
predicted count rate as a function of energy for the entire REXIS instrument \cite{2013...Allen}.
The collection of contaminants on the surface of the CCD is mitigated by the implementation of
pre-flight handling procedures, however the migration and accumulation of organic contaminants on
optical blocking filters and detectors consisting primarily of carbon has been observed in both the
Chandra X-ray observatory Advanced CCD Imaging Spectrometer (CXO-ACIS) \cite{2003SPIE.4851...89P},
the Suzaku XIS (c.f. \cite{2013...XISMON}) and many others.  Such contaminants are suspected to
originate from the spacecraft itself \cite{2013...Bautz} with an accumulation rate has been observed
to have a strong temperature dependence.  Therefore the final contamination properties will vary
from mission-to-mission depending upon the components, geometry and temperature of the spacecraft;
difficult to assess prior to launch.  
%Recent assessments by different teams for the Hinode XRT
%\cite{2008JSASS..56..536U} and the 
Based on the observed properties of the CXO-ACIS and Suzaku XIS contamination layers a
representative contamination layer consisting of C and O with a atomic ratio C/O=6 and total
thickness of 54 $\mu$g/cm$^2$ has been assumed for our simulations.  The minimum detection times for
elements key to the identification of Bennu outlined in \sref{\S}{section:asteroid_identification}
are given in \sref{table}{table:detection_times}.

\section{Meteorite Classification}\label{section:asteroid_identification}
\begin{figure}
	\centering
	\begin{subfigure}{0.49\textwidth}
		\includegraphics[width=\textwidth]{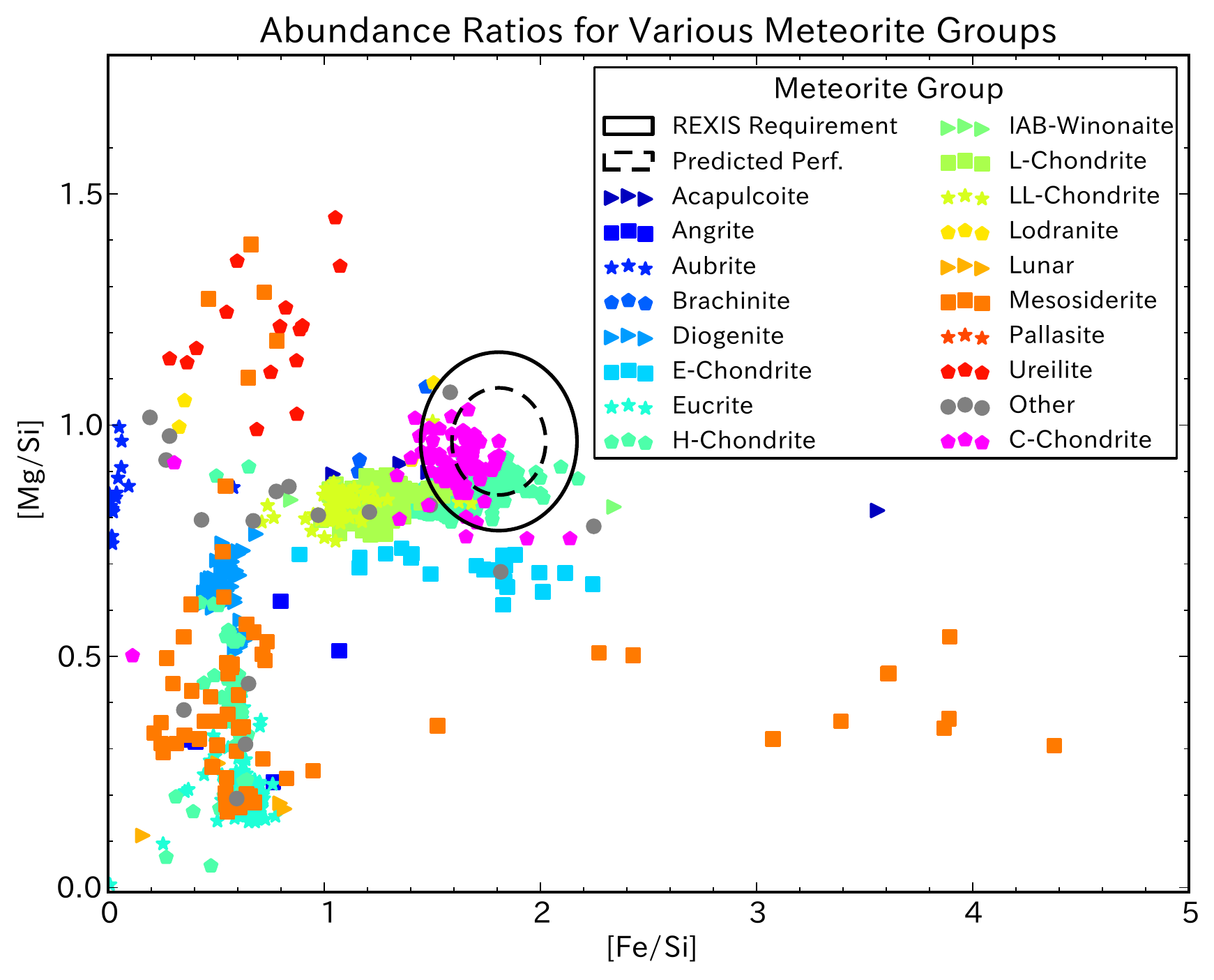}
		\caption{}%Identification of the meteorite group by [Mg/Si] vs. [Al/Si] mass ratios.}
		\label{figure:asteroid_class_id}
	\end{subfigure}
	\begin{subfigure}{0.49\textwidth}
		\includegraphics[width=\textwidth]{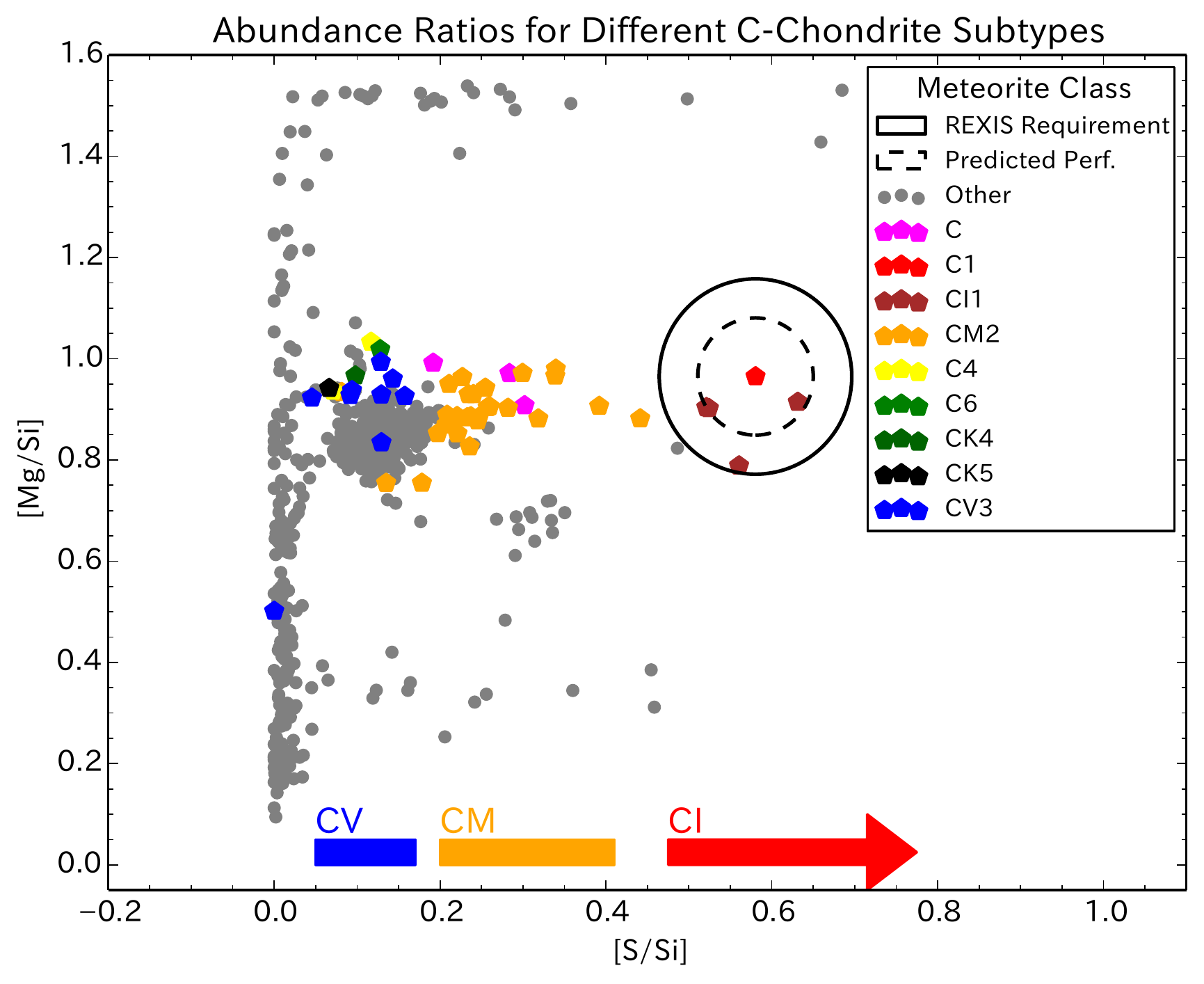}
		\caption{}%Asteroid subgroup identification by weight ratio: [Mg/Si] vs. [S/Si]}
		\label{figure:asteroid_subclass_id}
	\end{subfigure}
	\caption{Ground observations have shown Bennu to most closely resemble a C1 or CM chondritic
	meteorite.  Two of the primary objectives of the REXIS mission is to (a) determine whether or not
	Bennu is indeed a Chondrite, and to (b) determine which type of Chondrite based on observations
	of Sulfur.  The expected performance and requirement values are show in the above plots for REXIS
	as a dashed and solid line respectively.  The measurements of the elemental abundance ratios for
	the individual meteorites were carried out in the lab and the results are given in
	\cite{2004AMR....17..231N}.}
\end{figure}
The primary objective of the REXIS instrument is the identification 101955 Bennu elemental surface
composition through the observation of the Fe-L, Al-K, Mg-K, and Si-K complexes,  as well as the
S-K$\alpha$ and S-K$\beta$ fluorescence lines.  Using data from \cite{2004AMR....17..231N} it was
determined that the primary means of identification will be carried out through measurement of the
mass abundance ratios [Mg/Si] and [Fe/Si] (see \sref{figure }{figure:asteroid_class_id}) which will
enable REXIS to test previous observations of Bennu (see \sref{\S}{section:bennu}).  Abundance
ratios are typically used in place of simple element abundances as these are relatively insensitive
to changes in solar states and unexpected changes in instrument performance.  In order to
definitively classify Bennu as a carbonaceous chondrite REXIS is required to measure these abundance
ratios to within 20\% of their true value at a confidence level of 10$\sigma$; the current best
estimate for the expected performance of the instrument now stands as 18.8\% \cite{2013...Inamdar}.
%For [Al/Si]

REXIS is not only designed toward determining if Bennu is a carbonaceous chondrite, but will also be able
to distinguish the subgroup to which Bennu belongs.  To accomplish this the [Mg/Si] and [S/Si]
abundance ratios are required to be measured (see \sref{figure }{figure:asteroid_subclass_id}) to
within 20\% of their true value with an expected performance of 12\%.

\section{Coded Aperture Imaging Simulation and Reconstruction}\label{section:imaging}
\begin{figure}
	\centering
	\includegraphics[width=\textwidth]{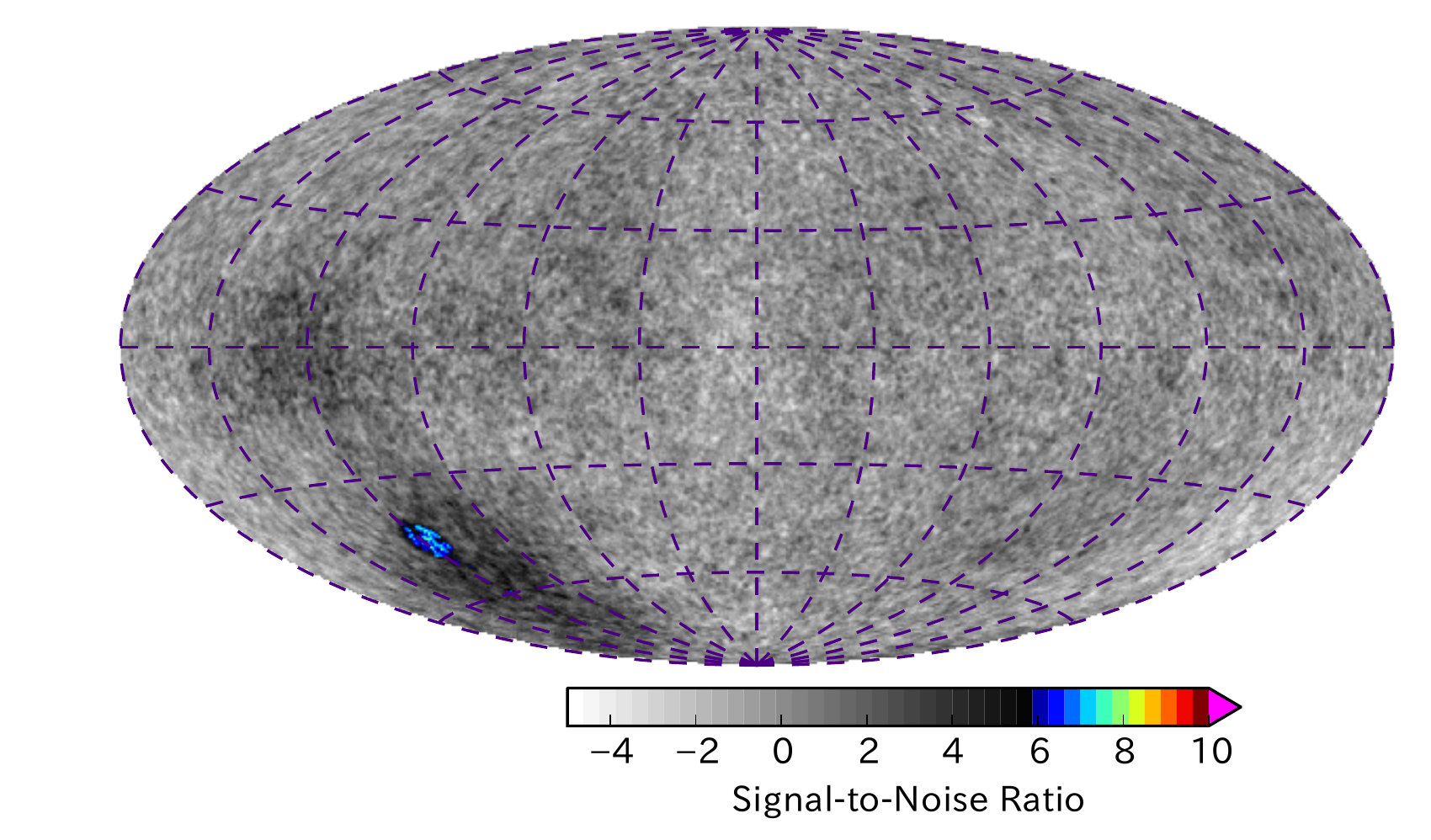}
	\caption{The complete reconstructed, simulated image of Bennu with a single 50 m scale source
	with a factor of 5 abundance enhancement observed with a total integration time of 20 days.  The
	peak signal-to-noise ratio is 7.8; a simple scaling indicates that a detection at the 6$\sigma$
	level is expected for factors of 3.75 enhancement in Fe abundance.}
	\label{figure:coded_aperture_image}
\end{figure}
An imaging simulation has been carried out for a fixed orbital radius of 1 km around a spherical
asteroid with a radius of 280 m in the Fe-L and Mg-K bands in order to assess the sensitivity of the
REXIS instrument to enhancements of elemental abundance on 50 m scales on the asteroid surface.
Initially a flux map of the asteroid is initially generated on a AITOFF-Hammer equal area projection
and regions of enhanced flux are added to the map simulating regions with enhanced elemental
abundance.

The imaging simulation proceeds in two step: first the collection of data in 600 second time steps
on the detector plane is simulated and a detector plane image generated at each step.  This is
coarser than the 4 second integration time of the REXIS detector plane (see
\sref{\S}{section:rexis}) in order to carry out the simulations within a short time frame:
approximately 48 hours for 20 days of simulated observation per CPU core.  Next the individual detector
plane images are used to reconstruct individual observation images which are then co-added over the
surface of the asteroid in a AITOFF-Hammer projection.  A search over this final map is then
conducted for statistically significant flux enhancement.

The data collection simulation is carried out by initially calculating the sub-solar point and the
position of the OSIRIS-REx spacecraft at each step as a function of time fully taking into account
the orbit of OSIRIS-REx as well as the rotation of 101955 Bennu.  At each simulation step the solar
flux incident at each point on the asteroid is first calculated which is then used to calculate the
intensity of the X-ray emission from the surface of the asteroid to the REXIS instrument.  The
intensity map is then re-projected into the instrument's local field of view on a tangential
projection and convolved with the coded aperture mask using a fast Fourier transform (FFT) producing
an incident intensity map on the REXIS detector plane.  This map is then filtered according to the
active area of the CCD and the expectation value of the number of counts in each CCD super-pixel
calculated.  A simulated observation is then carried out by sampling each pixel on a Poisson
distribution using the expectation value producing a single detector plane image.  The detector
plane image together with the spacecraft position and sub-solar position are stored to disk at each
step.

The image reconstruction process is carried out first by reconstructing the sky image from the
stored detector plane image through an FFT convolution with the coded-aperture mask using a balanced
correlation where open mask pixels are assigned a value of 1 and closed mask pixels are assigned a
value of $\rho/(\rho-1)$ where $\rho$ is the open hole fraction \cite{1978ApOpt..17..337F}; mask
pixels outside of the circular mask pattern area are removed from the reconstruction by setting
their values to 0.  Similarly the sum of all events over all pixels in the detector plane active
area is set to 0 by subtraction the mean number of events taken over all active elements.  In
carrying out this procedure an automatic background subtraction of the reconstructed sky image is
carried out giving an excess map in sky coordinates.  Using the OSIRIS-REx spacecraft position the
sky image is reprojected to asteroid coordinates on an AITOFF-Hammer equal area projection.  The
equal area projection is used in order to carry out the subsequent analysis on equal sized surface
area elements on the asteroid surface as well as to co-add images in a coordinate space that is
independent of the spacecraft view factors to the asteroid surface.

In addition a background estimation procedure is also carried out over the asteroid surface by
random re-assignment of the positions for individual sky images before re-projection within a fixed
time interval around individual images.  This time-shuffling procedure has been introduced in order
to remove the large-scale anisotropies associated viewing the bright limb of the asteroid and the
attendant effect within the co-added sky images in asteroid coordinates.  The same goal can be
achieved in the detector image space by co-adding detector plane images over a rolling time window
to produce a background detector map which can then be subtracted from the individual detector plane
before reconstruction of the sky image (see \sref{figure }{figure:coded_aperture_image}).  Although
the variability of the solar flux has not yet been added to this imaging simulation the purpose of a
subtracting images at nearby times is to test the procedure that will be required when dealing with
long-term variability of the solar state over the REXIS observation period.  In the final analysis
procedure short-term variability will be handled using input from the SXM to sort individual images
into separate solar states for co-addition into separate asteroid maps for each state in order to
avoid the appearance of overabundant regions due only to enhanced fluxes caused by solar flares in
individual images.

Initially the imaging simulation was carried out assuming a 4 MK coronal temperature with an
emission measure of $10^{44}$ cm$^{-2}$, however it was subsequently determined that the flux
enhancements required for the detection of a source on the surface of a C1 Chondrite was unphysical.
It was determined shortly thereafter that enhancements Fe and Mg elemental abundance under exposure
from a 6 MK sun would yield detectable sources on 50 m scales.  Since approximately 95\% of
observations are likely to be carried under exposure from a $T\geq6$ MK sun (see \sref{figure
}{figure:goes_em_temp} and \sref{\S}{section:asteroid_spectrum}) assuming the same conditions as the
previous solar minimum: equivalent to 19 days of observation, however intervals of time during which
the short wavelength GOES X-ray monitors (0.5-4 \AA) were unable to measure significant fluxes from
the sun at low solar states were excluded.  Assuming there periods represent lower coronal
temperatures the expected REXIS observing time during 6 MK solar state requires a downward
correction of up to 50\%.  A more careful evaluation of this is currently underway
\cite{2013b...Allen}.

\section{Collimator Mode Simulation and Reconstruction}\label{section:recon}
\begin{figure}
	\centering
	\begin{subfigure}{0.49\textwidth}
		\includegraphics[width=\textwidth]{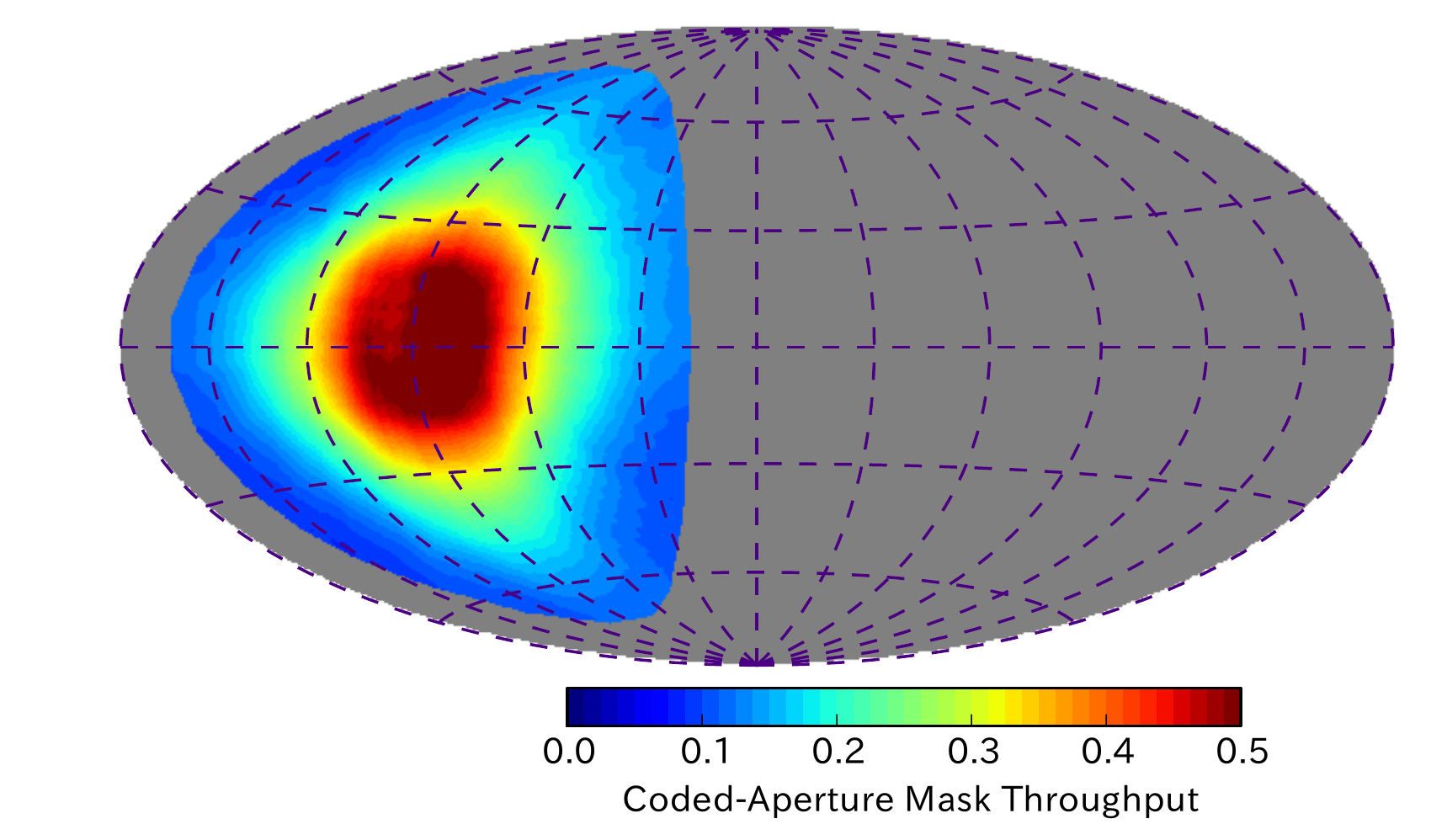}
		\caption{The REXIS Instantaneous Response Function}
		\label{figure:open_fraction}
	\end{subfigure}
	\begin{subfigure}{0.49\textwidth}
		\includegraphics[width=\textwidth]{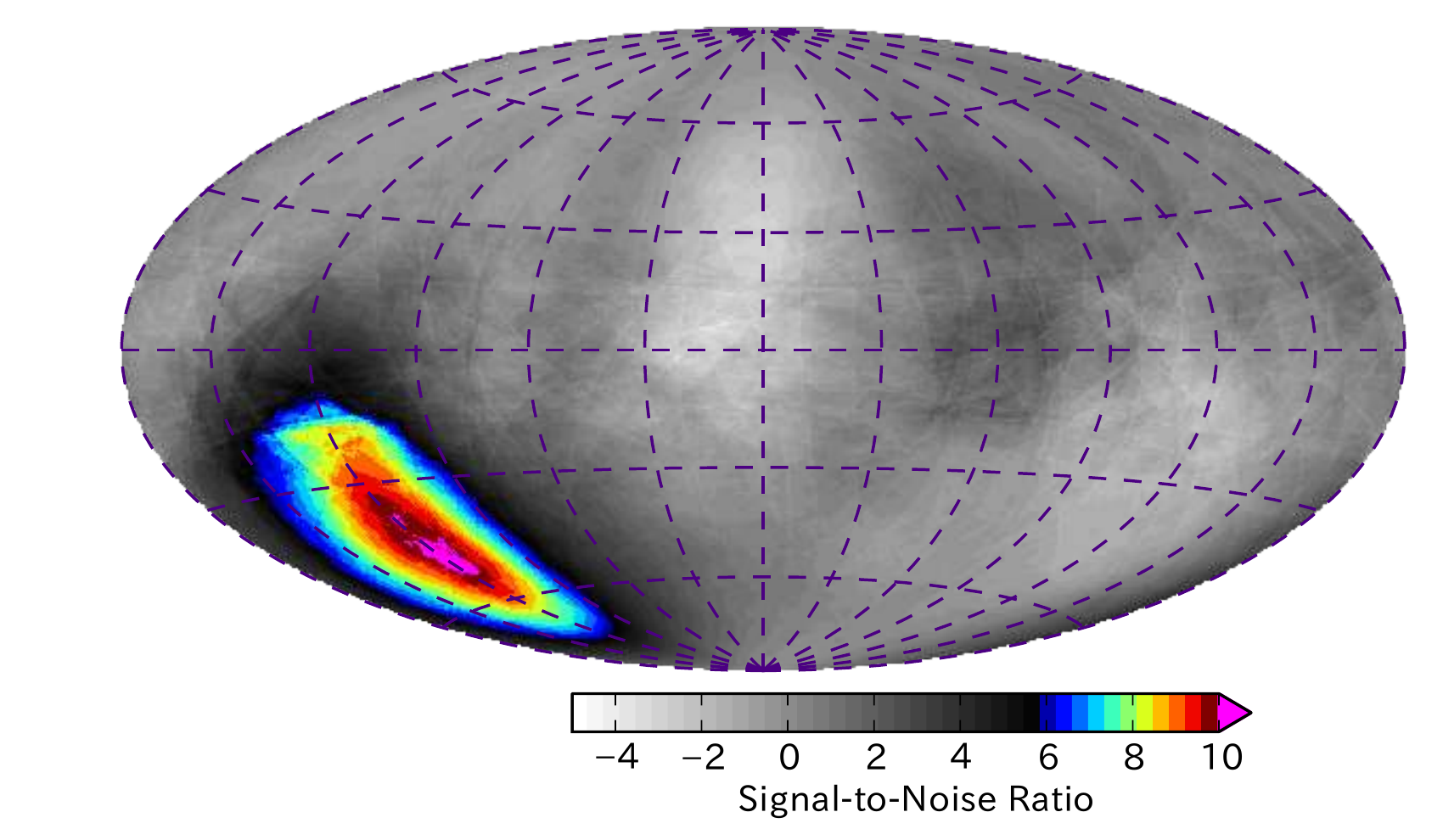}
		\caption{Reconstructed Collimator Mode Image}
		\label{figure:collimator_reconstruction}
	\end{subfigure}
	
	\caption{(a) The response function of the REXIS instrument is defined by the coded aperture mask
	throughput shown for REXIS during nadir pointing at $(lat,lon)=(-90.0,0.0)$.  For reconstruction
	of elemental abundance maps at low energies the response function can be further reduced by
	weighting the exposed areas by the intensity of the solar exposure. (b) The reconstructed image
	of elemental abundance enhancements for a single Fe enhancement located on the surface of 101955
	Bennu identical to that detected in \sref{figure }{figure:coded_aperture_image}. The peak signal
	to noise of the detected source 10.2$\sigma$.}
	\label{figure:collimator_mode}
\end{figure}

The simulation and reconstruction of images in collimator mode is carried out in a similar manner to
that of the imaging simulation and reconstruction described in \sref{\S}{section:imaging}.  The key
difference is that the response function of the entire REXIS instrument (see \sref{figure
}{figure:open_fraction}) is used rather than individual mask pixels for the reconstruction of
individual asteroid images.  This leads to degraded angular resolution but somewhat higher
sensitivity to larger scale elemental abundance enhancements on the asteroid surface.  Additionally
the terminator orbit configuration for the REXIS observations also proves to be advantageous,
particularly for lower energy elements where the relative contribution of the CXB induced X-ray
fluorescence is low and the emission can be approximately considered to originate from one half of
the REXIS field of view.

In order to simulate the collimator mode reconstruction the same simulated data set consisting of
detector plane images, the OSIRIS-REx spacecraft position, and sub-solar position as a function of
time is generated as described in \sref{\S}{section:imaging}.  The position of each detector plane
image is then used to project the collimator response function onto the surface of the asteroid
weighted by the total number of events registered in the detector plane over the energy band of
interest.  For low energy elements whose fluorescence lines are primarily induced by solar X-ray the
emission can be considered to have originated from the illuminated half of the asteroid.  Weighting
of the response function by the incident solar exposure, effectively cutting the response function
in half, improves the collimator mode spatial resolution by approximately a factor of two.

A similar method for the estimation of a background map is also carried out by time shuffling
collimator and co-adding images on the asteroid surface as was described in
\sref{\S}{section:imaging}.  As before a time window of 12000 seconds, equivalent to 20 exposures,
was adopted.  After generation the raw asteroid map is subtracted from the background asteroid map
giving excess as a function of position with a spatial resolution on the asteroid surface of 250 m.

Studies are underway to further characterize the performance of this reconstruction technique for
the REXIS instrument as a function of source strength and extension \cite{2013b...Allen}, however
low spatial resolution detections of source on the asteroid surface have been simulated resulting in
high signal-to-noise detections for Fe sources on scales of 50 m and larger with a factor of 5
element abundance enhancement (c.f.  \sref{figure }{figure:collimator_reconstruction}).

\section{Optimization of the REXIS Imaging System}\label{section:optimization}
The REXIS instrument (see \sref{figure }{figure:rexis}) coded aperture mask and imaging system has
been optimized using the imaging simulation described in \sref{\S}{section:imaging}.  The primary
parameters for optimization were the radius of the mask pattern, the mask open hole fraction, and
the pixel pitch of the mask as well as that of the detector plane; see \sref{table }{table:rexis} for
a list of the current parameters.

The optimization procedure was as follows: the input parameters for a candidate mask are set, i.e.
pattern radius, pixel pitch, support grid width, and open hole fraction, each  mask pattern was
optimized through generation of 1000 candidate masks.  Next a extreme high flux point source
simulation using the REXIS detector plane with a pixel pitch of 0.368 mm and iterating the source
over the entire REXIS field of view in order to measure the maximum achievable signal-to-noise
ratio.  From the sample of 1000 candidate masks the mask showing the best performance over the
entire field of view is chosen for use in the subsequent simulations discussed below.

The first parameter considered for optimization was the mask and detector plane pixel pitch.  A
series of simulations for the full observation period of REXIS was carried out for a test detector
with pixel pitch of 0.368 mm and a mask pixel pitch of 1.152 mm with a region of enhancement
identical to that described in \sref{\S}{section:imaging}.  This leads to an improvement in the
angular resolution from $26.2^\prime$ to $20.7^\prime$ at the cost of a 7\% decrease in the total
count rate resulting from a decrease in the mask throughput.  This is due to the  chemical etching
process used for the production of the coded aperture mask which requires that the support grids
running between individual elements is are at least equal to the thickness of the material; in this
case 100 $\mu$m.  The on-axis throughput of a coded aperture mask with support grids can be
calculated by $T^\prime= T(1-g/m_p)^2$, where $T$ is the throughput expected for a perfect mask
without support grids, i.e.  the open hole fraction, $m_p$ is the mask pixel pitch and $g$ is the
width of the support grid.  For the test carried out here a 4.8\% decrease in the on-axis
throughput, from 0.437 to 0.417, occurs in the transition from the 1.536 mm mask to the 1.152 mm
mask.  Although this will not have a significant negative impact on our observations the 1.536 mm
mask was retained since a significant positive impact was also not observed.  From a fabrication and
integration perspective the tightening of the alignment requirements for the use of the finer pitch
mask was not warranted given the lack of any clear advantage.

It was recognized early on that the performance of the spectral observations could be improved by
increasing the throughput of the coded aperture mask at the expense of imaging sensitivity.  In
order to ascertain the effects on imaging the same simulation was run using coded aperture masks
with ideal open fractions of 0.25, 0.35, 0.4, 0.6, 0.65, and 0.75.  A series of 10 simulations were
run for each mask pattern in order to measure not only the peak signal-to-noise ratio of each source
but also to check the fluctuation level of the result.  No significant detections were found in any
of these test runs.  Additional simulation were initiated for masks with open fraction of 0.45 and
0.55 in a search for potential optimization of the image reconstruction unfortunately the result
here was also negative, confirming the open hole fraction of 0.5 to be optimal for the observation
of Bennu.

\section{Discussion}\label{section:discussion}
The REXIS instrument has been optimized for observation of the near-Earth asteroid 101955 Bennu as
part of the OSIRIS-REx mission and seeks to identify a meteorite classification of Bennu through
measurement of elemental abundances by observation of solar X-ray and CXB induced fluorescence lines
emitted from the surface of the asteroid.  REXIS will also search for localized areas of enhanced
elemental abundance with scale sizes of approximately 50 m in order to aid in the selection of a
sample collection site using a coded aperture mask capable of resolving features as small as 5.6 m
on the surface of the asteroid.  REXIS will be the first instrument to employ this technique for use
on a planetary mission and owing to its large field of view will also maintain similar capabilities
to instruments which have previously been used in this capacity.

\acknowledgements{This work was supported under NASA Grants NNX12AG65G and NNG12FD70C.  We would
like to thank Martin Elvis for making the initial contact between the Harvard College Observatory
and Massachusetts Institute of Technology groups which have participated in the development of
REXIS, Lucy Lim for providing invaluable advice on the utility of solar X-ray monitors, George
Sondecker for his work during the proposal and early phase of this project, as well as Kevin Ryu and
Vsyhi Suntharalingam for the preparation of the REXIS CCDs.
}

%REFERENCES
\bibliographystyle{spiebib}
\bibliography{REXIS.8840-24.bbl}
\end{document}